\documentclass[12pt]{article}
\begin{document}
\title{{\itshape Rejoinder on ``Conjectures on exact solution of
three-dimensional (3D) simple orthorhombic Ising lattices"}%
\footnote{Supported by NSF grantPHY 07-58139}}

\author{Jacques H.H. Perk%
\footnote{Email: perk@okstate.edu}\\ \cr
145 Physical Sciences, Oklahoma State University,\\
Stillwater, OK 74078-3072, USA\footnote{Permanent address}\\
and\\
Department of Theoretical Physics, (RSPE), and\\
Centre for Mathematics and its Applications (CMA),\\
Australian National University,\\
Canberra, ACT 2600, Australia}

\maketitle

\begin{abstract}
It is shown that the arguments in the reply of Z.-D. Zhang defending his
conjectures are invalid. His conjectures have been thoroughly disproved.
\end{abstract}

After all the discussion about his paper \cite{zdz,wmfc,zdz2,wmfc2,jhhp},
Zhang seems to have only one real issue left with the two comments
\cite{wmfc,jhhp}, still wrongly believing \cite{zdz3} that the
free energy of the three-dimensional Ising model is not analytic
at $\beta\equiv1/(k_{\rm B}T)=0$, $H=0$. His further arguments are
irrelevant or dealt with adequately in \cite{wmfc,jhhp}.

His objection that Gallavotti and Miracle-Sol\'e set $\beta=1$ is
not valid. One often uses dimensionless parameters $K_i=\beta J_i$,
($i=1,2,3$), $h=\beta H$. Equivalently, one can absorb the $\beta$
into the coupling constants, setting $\beta=1$, so that $K_i\equiv J_i$,
$h\equiv H$. Infinite temperature is then the
limit $K_i\to0$, $h\to0$, such that all ratios are kept fixed.%
\footnote{The reduced free energy per site $\beta f$  is often rewritten
$\beta f=\phi(\{K_i\},h)=\phi(\{\beta J_i\},\beta H)$
with some function $\phi$. Setting $\beta=1$ is no loss of generality,
as one can easily restore the $\beta$-dependence by the replacements
$J_i\to\beta J_i$, $H\to \beta H$, $f\to\beta f$.}

His point on  \cite{ly} is also not valid. Writing
$z\equiv\exp(-2\beta H)$, see eq.\ (23) of \cite{ly},
and keeping $\beta H$ fixed in the limit $\beta\to0$,
the Ising model partition function on an arbitrary lattice
with $N$ sites becomes $Z=(z^{1/2}+z^{-1/2})^N,$
so that all infinite-temperature zeros of $Z$ occur
at $z=-1$, i.e.\ for the purely imaginary magnetic field \cite{ly}
$H=\pm i\pi k_{\rm B}T/2=\pm i\infty$. There is no
$T=\infty$ singularity at $H=0$, $z=1$.%
\footnote{In Zhang's paper \cite{zdz}, $H\equiv0$, $z\equiv1$,
$Z=2^N$ for any lattice with $N$ sites at $T=\infty$, which is far
from $z=-1$. Hence, the infinite-system dimensionless free energy
$\beta f$ is analytic at $\beta=0$ also by the general theory of
Yang and Lee \cite{yl}.}

\newpage

\section*{Added Comments}

It has saddened my heart to see arXiv:0812.0194v3 and to see that
Zhang is still not convinced that his conjectured results are in error.
For his benefit and the benefit of others who might be confused by his
remarks I am adding the following to show why his newer statements
are also in error.

The probability interpretation of statistical mechanics
involves $Z$ and thus $\beta f$. Therefore, that $f\!\to\!\infty$
for the free energy per site as $T\!\to\!\infty$ is of no
physical significance. Series and
analyticity determinations must be starting from the reduced
free energy $\beta f=f/k_{\rm B}T$, not $f$, near $T=\infty$.

Also, as long as (the real part of) $f$ is negative, (which
is easily checked for the Ising model at arbitrary
temperature), $Z=\exp(-N\beta f)$ blows up and $1/Z\to0$
as $N\to\infty$. So $\lim_{N\to\infty}1/Z=0$
occurs for all finite temperatures even at the critical
point where the Yang--Lee zeroes pinch the
real axis. It is a misinterpretation of the Yang--Lee
papers to study zeroes of $1/Z$. The Boltzmann probability
is given by $\exp(-\beta{\cal E})/Z$, with $Z$ a polynomial
in $z=\exp(-2\beta H)$, $H$ being the scaled magnetic field.
This probability can only show anomalous behavior if zeroes of
$Z$ approach as $N\to\infty$, as is clearly explained
by Yang and Lee in their two famous 1952 papers.

Finally, there are many papers proving analyticity of the
reduced free energy in $1/T$ around $T=\infty$, for both
classical and quantum models that generalize the Ising model. Such
proofs have been published by many groups in Europe, Russia,
Japan and America. I have only quoted a few mostly Ising references
in my comment. The exactness of the series is rigorously proved
many times. It is also backed up by numerical work, as Pad\'e
analysis of the series expansion is in excellent agreement with
other numerical work, such as Monte Carlo calculations. There
can be no such evidence backing up Zhang's work, as it violates
rigorously established theorems.

\section*{Added Comments 2}

As Zhang is stacking up further errors in his fourth version
arXiv:0812.0194v4 and repeats his errors to the followers of
his blog in China, I feel that I have little choice than adding
another reply.

In the first place, under ``solving the three-dimensional Ising model"
is understood the calculation of some basic thermodynamic quantities
using standard equilibrium statistical mechanics starting from
$\hat H$ in [1], i.e.\ the
usual Ising interaction energy. Any talk about time dependence, ergodicity,
relativity, quantum mechanics, black holes, etc., is a distraction
from the well-defined problem and changing it to a different problem.

Standard equilibrium statistical mechanics states that the free
energy per site $f$ is calculated using
\begin{equation}
\beta f=\frac1N\beta F=-\frac1N \log Z,
\end{equation}
with $Z$ the partition function and $N$ the number of sites. It is
$\beta f$ which is calculated first and it is $\beta f$  for which series expansions are done. That $f$ becomes $\infty$ at $T=\infty$ is of no significance, as is obvious from one of the basic formulae in thermodynamics, $F=U-TS$. This says that at $T=0$ we have $F=U$ the
(internal) energy, while at $T=\infty$ we find $F/T=-S$, i.e.\ minus the
entropy. This explains again that $F/T$, or equivalently $\beta F$, is
preferred over $F$ at high temperatures.

Next, consider eq.~(1) in Zhang's paper [1], and define
\begin{equation}
C=\max\{|J|,|J'|,|J''|\},
\end{equation}
the absolute maximum of the coupling constants. Then it is easily
checked that
\begin{equation}
|\hat H|\le 3NC,\quad |Z|\le2^N{\rm e}^{3NC},\quad
\frac1N\log|Z|\le\log2+3C,
\end{equation}
as the number of terms in $Z$ is $2^N$. The only way to get singularities
in $\beta f$ is the occurrence of zeroes of $Z$, as explained excellently
by Yang and Lee, whom Zhang chooses to misquote. Zhang's work is in stark
contradiction with the work of Yang and Lee as a result.

Finally, using the logic of his conjecture we can solve almost any
problem in statistical mechanics, but with incorrect results. The idea
of going to higher dimension is a generalization of the fundamental
theorem of calculus $\int_a^b f'(x){\rm d}x=f(b)-f(a)$, with
examples the theorems of Green, Stokes, Gauss, and the Wess--Zumino term.
For the ``guess" to work, Zhang must show that the integrand
in four dimensions is a derivative of the integrand in 3D (or a
discrete version with integrand replaced by summand and derivative by
some kind of difference). In [1] I cannot find that Zhang understands this
and, as the conjecture is rigorously proved to be false, it makes
no sense to further discuss this matter here.

\section*{Added Comments 3}

Most proofs of the analyticity of free energies and correlation functions
use linear correlation identities of Schwinger--Dyson type, known under
such names as the BBGKY hierarchy, Mayer--Montroll or Kirkwood--Salzburg
equations. If Zhang had read \cite{lp} in detail and understood, he would
have found more than one way to arrive at a proof of analyticity of
$\beta f$ at $\beta=0$. He would also have found \cite{gmr} (ref.~[12]
cited in \cite{lp}). Let me try to explain the proof in more
down-to-earth terms using an identity of Suzuki \cite{suz1,suz2},
restricted to the isotropic Ising model on a simple cubic lattice with
periodic boundary conditions and of arbitrary size, i.e.
\begin{equation}
\bigg\langle\prod_{i=1}^n\sigma_{j_i}\bigg\rangle=\frac1n\sum_{k=1}^n
\bigg\langle\bigg(\prod_{i=1\atop i\ne k}^n\sigma_{j_i}\bigg)
\tanh\bigg(\beta J\sum_{l\;{\rm nn\;of}\;j_k}\sigma_{l}\bigg)\bigg\rangle,
\label{suzid}\end{equation}
where $j_1,\ldots,j_n$ are the labels of $n$ spins and $l$ runs through
the labels of the six spins that are nearest neighbors of $\sigma_{j_k}$.
The isotropy assumption, i.e. $J_1=J_2=J_3=J$, is made to simplify the argument, but can be easily lifted. Averaging over $k$ has been added
in (\ref{suzid}), so that all spins are treated equally.

Next we use
\begin{equation}
\tanh\bigg(\beta J\sum_{l=1}^6\sigma_l\bigg)=
a_1\sum_{(6)}\sigma_l
+a_3\sum_{(20)}\sigma_{l_1}\sigma_{l_2}\sigma_{l_3}+a_5
\sum_{(6)}\sigma_{l_1}\sigma_{l_2}\sigma_{l_3}\sigma_{l_4}\sigma_{l_5},
\label{suzidt}\end{equation}
where the sums are over the 6, 20, or 6 choices of choosing 1, 3, or 5
spins from the given $\sigma_1,\ldots,\sigma_6$. It is easy to check that
the coefficients $a_i$ are
\begin{eqnarray}
&&a_1=\frac{t(1+16t^2+46t^4+16t^6+t^8)}{(1+t^2)(1+6t^2+t^4)(1+14t^2+t^4)},
\quad a_3=\frac{-2t^3}{(1+t^2)(1+14t^2+t^4)},\nonumber\\
&&a_5=\frac{16t^5}{(1+t^2)(1+6t^2+t^4)(1+14t^2+t^4)},
\quad t\equiv\tanh(\beta J).
\label{suzida}\end{eqnarray}
The poles of the $a_i$ are at $t=\pm{\rm i}$, $t=\pm(\sqrt{2}\pm1){\rm i}$,
and $t=\pm(\sqrt{3}\pm2){\rm i}$. It can also be verified, e.g. expanding
the $a_i$ in partial fractions, that the series expansions of the $a_i$ in
terms of the odd powers of $t$ alternate in sign and converge absolutely
as long as $|\beta J|<\arctan(2-\sqrt{3})=\pi/12$.

The system of equations (\ref{suzid})--(\ref{suzida}) can be viewed as
a linear operator on the vector space of all correlation functions of
the 3d Ising model. It is easy to estimate the norm of this operator,
from the $32n$ terms in the right-hand side (RHS) of (\ref{suzid}) after
applying (\ref{suzidt}), using the alternating sign property of the
$a_i$'s. It follows that we only need to study
\begin{equation}
6a_1+20a_3+6a_5=\frac{2t(t^2+3)(3t^2+1)}{(1+t^2)(1+14t^2+t^4)}
\label{suzidi}\end{equation}
for purely imaginary $t$ to find the desired upper bound $r$ for the
norm. We then have that the RHS of (\ref{suzid}) is bounded by $rM$,
where $M=\max|\langle\sigma\cdots\sigma\rangle|$ with the maximum taken
over all $32n$ pair correlations in the RHS. Obviously, $M\le1$ if
$\beta\ge0$ and real. We must stress that the bound is also valid if all
the $a_i$'s are replaced by the power series in $t$ obtained from the
power series of the $a_i$'s replacing each term by its absolute value.

According to the above we set $r$ equal to the absolute value of (\ref{suzidi}) for imaginary $t$. We can then show that $r<1$ for
\begin{eqnarray}
&&|t|<(\sqrt{3}-\sqrt{2})(\sqrt{2}-1)=0.131652497\cdots,
\quad\hbox{or}\nonumber\\
&&|\beta J|<\arctan[(\sqrt{3}-\sqrt{2})(\sqrt{2}-1)]=0.130899693\cdots.
\label{suzidn}\end{eqnarray}

To prove analyticity of $\beta f$ in terms of $\beta$ at $\beta=0$ it
suffices to study the internal energy per site or the nearest-neighbor
pair correlation function, as
\begin{equation}
u=\frac{\partial(\beta f)}{\partial\beta}=
-3J\langle\sigma_{000}\sigma_{100}\rangle.
\end{equation}
We apply (\ref{suzid}) to $\langle\sigma_{000}\sigma_{100}\rangle$,
then we apply (\ref{suzid}) on each of the new correlations, and we
keep repeating this process. We may now and then encounter a correlation
with zero $\sigma$ factors $\langle1\rangle=1$, at which the process ends.
As each other correlation vanishes with a power of $t$, we then
generate the high-temperature power series to higher and higher orders,
for arbitrary given size of the system. The series is absolutely
convergent as the sum of the absolute values is bounded by
$\sum r^j<\infty$ when (\ref{suzidn}) holds. We can at first assume
that $\beta\ge0$ and real. But from the absolute convergence we
also obtain a finite radius of convergence in the complex $t$ and $\beta$
planes.

Increasing the size of the system, we conclude that more and more terms
become independent of the size, whereas the remainder rapidly tends to
zero. Therefore, the series converges to the thermodynamic limit, and we
have once again proved that the free energy and all correlation functions
are analytic in $t$ and in $\beta$, as long as (\ref{suzidn}) holds. This
bound is a rough estimate and much better ones have been given in the
literature. Furthermore, adding a small magnetic field $H$ and generalizing the
steps in the above, we can also conclude that all correlation functions
are finite for small enough $\beta$ and $H$, so that there are no
Yang--Lee zeroes near the $H=0$ axis for small $\beta$.

In conclusion once more, papers \cite{zdz,zdz2,zdz3} contain serious errors and are beyond repair.

\end{document}